\documentstyle[12pt]{article}
     \begin{document}
      \vskip 2. true cm
      \begin{center}
      {\large \bf Considerations on localization
      of macroscopic bodies} \\

      \vskip 1.4 true cm
        B. Carazza \\
       \vskip .1 true cm
\par 
       {\it Dipartimento di Fisica dell' Universit\`a, 
       viale delle Scienze, I43100 Parma, Italy}\\
        {\it INFN Sezione di Cagliari, Italy}\\  
      \end{center}
\par \vskip 2mm
\begin{center} {\bf  Abstract} \end{center}
 \vskip 1mm
\begin{quote}

Position holds a very special role in understanding the classical 
behaviour of macroscopic bodies on the basis of quantum 
principles. This lead us to examine the localised states of a large 
condensed object in the context of a realistic model. Following the 
argument that an isolated macroscopic body is usually described 
by a linear superposition of low-lying energy eigenstates, it has 
been found that localised states of this type correspond to a nearly 
minimum-uncertainty state for the center of mass.
An indication is also given of the dependence of the
center of mass position spread on the number of
constituent particles.
 This paper is not offered as an answer to the 
intriguing question of the preferred role played by the position 
basis, but will hopefully provide some contribution to the 
quantum modelling of multi-particle systems.

\end{quote}
\vskip 1.5 true cm
Key words: Macroscopic states, spatial macrosuperpositions,
preferred basis, approximate position  eigenstates.

 \newpage
      \noindent{\bf 1. INTRODUCTION}
      \vskip .6 true cm
 \par
Classical physics has explained a variety of phenomena, ranging 
from sound to heat, in terms of systems of moving particles. These 
achievements are firmly based on the primary notions of position 
and trajectory first deduced from the observation of common 
everyday objects. It was possible to do this because macroscopic 
bodies appear to be fairly accurately localised, which means that 
position undoubtedly plays a privileged role in our knowledge of 
the external world. However, if the problem is approached in 
terms of quantum physics, disregarding Bohr's point of view and 
applying it to all objects (not only microscopic ones), a number of 
difficulties are encountered.
 \par
 We could assume, when looking at the moon, that we see it in a 
certain position because it is described by a wave packet that is 
well localised precisely there, and not because our glance caused the 
moon's wave packet to collapse into that state.
Generally speaking, it may be wondered why spatial 
macrosuperpositions have not been encountered, even though 
they are theoretically possible, whereas macroscopic objects do 
tend to be found in (approximate) eigenstates of position.
 \par
All this concerns the general debate about the classical limits of 
quantum formalism. Recent work on the subject has been 
successful thanks to the use, for instance, of the functional 
representation of Hilbert space in terms of coherent states 
\cite{kla}, the 
Wigner function \cite{wig} and, above all, such concepts as decoherence 
\cite{omn}. Many researchers believe that decoherence analysis also 
adequately explains why position is so important.
 \par
The best result we can hope for is to demonstrate that, within 
the macroscopic limit, a body is usually described by a statistical 
mixture, each element of the ensemble being associated with a pure 
state corresponding, for all practical purposes, to a well defined 
position of its center of mass. Efforts in this direction have reached 
their objective not only through application of the decoherence 
mechanism but also by using the model of spontaneous dynamical 
reduction proposed by Ghirardi, Rimini and Weber \cite{ghi}, later
refined as the Continuous Spontaneous Localisation model (CSL) 
\cite{pea}. The CSL model requires selection of the so-called preferred 
basis- i.e. the eigenmanifold in which reduction occurs - and the 
authors attribute this privileged role to position.
 \par
In conclusion, daily experience, common sense and theoretical 
studies all attribute preferred status to the position of a 
macroscopic object. The present paper is intended as a supplement 
to the debate about this intriguing role, a debate which also 
extends into the realms of philosophy. Our intention  
is not to explain why large objects tend to be found in 
approximate eigenstates of position but, since well localised states 
of macroscopic bodies evidently do exist, the more modest 
purpose is to examine their quantum properties and discover 
whether they normally possess certain characteristics.
 This will be applied to the case of an isolated 
macroscopic object, going beyond such familiar models as coupled 
harmonic oscillators \cite{cin}.
 \vskip .9 true cm

 \noindent{\bf 2. THE NATURAL STATES OF A MACROSCOPIC BODY}
 \vskip .6 true cm
 \par

Let us consider a large isolated object.
We can attempt to answer the following question: in what quantum 
mechanical state can an isolated macroscopic body ordinarily be 
found?
Although interaction of the system with its surroundings is so 
small that it can be considered isolated, it must be admitted -
with Landau and others \cite{lan}- that a macroscopic body can, in fact, 
never be in a strictly stationary state, due to the extraordinary 
density of its energy levels and to the action of surrounding 
objects, even if this tends towards zero. In other words, 
macroscopic objects are never at absolute zero temperature.
Yet their temperatures must be much lower the typical atomic 
binding energies which hold them together, in order to ensure the 
permanence and stability usually associated with a "body".
 \par
If macroscopic objects are in contact with their environment, 
they inevitably undergo the apparent GRW localisation caused by 
decoherence. This apparent collapse of the wave function may be 
caused, for example, by random scattering \cite{teg}.
However, the effects of the external environment on dynamics 
are not considered here. Since we wish to identify the type of pure 
state in which the macroscopic object is generally found, it will 
only be assumed, for the reasons stated above, that it is a linear 
superposition of low-lying eigenstates of total energy in a 
reference frame where average total momentum is zero. This is a 
means of taking into account the external environment at zeroth 
order, while the specification about the reference frame is 
necessary in that the total energy is not a Galilean invariant. 
 Such linear superpositions will be called "natural states".
 \par
 This manner of proceeding allows the mixing of internal energy
 levels with the kinetic energy of the center of mass.
 In the present case what is considered as a difficulty
 within the nuclear shell model \cite{gra} is actually welcome.
Let us now look at the properties of localised macroscopic 
systems with average energies slightly higher than the ground 
state. This will be done with reference to a realistic model.
 \vskip .9 true cm

 \noindent{\bf 3. THE MODEL OF A MACROSCOPIC BODY}
 \vskip .6 true cm
 \par
Let us now consider a system of $N$ spinless identical bosons with 
mass $\mu$ interacting in pairs through a two-body potential of the 
Van der Waals type - i.e., with a long-range attractive part and a 
short-range repulsive one. This corresponds to a realistic picture 
for an aggregation of noble gas atoms, as long as all of them are in 
their ground state and can thus be considered constituent 
elements. It is impossible to solve the eigenvalue problem for such 
a system in order to investigate a generic superposition of low-
lying energy states. However, we may proceed by applying the 
variational method to determine the parameters of a trial 
wave function, requiring that the average value of 
the Hamiltonian should be minimum. The possibility that 
this trial wave function is the ground state wave vector or 
corresponds exactly to a stationary state is minimal. It will 
basically be a linear superpositions of stationary states, the lower 
ones of which will presumably dominate.
 \par
This can now be used to calculate the spread of
 the center of mass position.
The trial state vector considered is the normalised and completely
 symmetrised state constructed from the product:
      \begin{equation}\label{e1}
      \prod^N_{i=1} \, \varphi_i(\mbox{\boldmath $r$}_i)
      \end{equation}
where each function  $\varphi_i(\mbox{\boldmath $r$}_i)$
  is centred on the $i^{th}$ of the $N$ 
contiguous lattice points  $\mbox{\boldmath $x$}_i$
 of a portion of regular lattice 
surrounded by a spherical surface whose origin coincides with the 
origin of the coordinates. Apart from this, all of the
functions $\varphi_i(\mbox{\boldmath $r$}_i)$
 have the same exponential-type form:
      \begin{equation}\label{e2}
      \varphi_i(\mbox{\boldmath $r$}) = \left \{ \begin{array}{l}
        D \, e^{-\lambda/2 |\mbox{\boldmath $x$}_i-\mbox{\boldmath $r$}|} 
        \quad \mbox{for $ |\mbox{\boldmath $x$}_i-
      \mbox{\boldmath $r$}| \leq a $}  \\
         0  \qquad \qquad \qquad \qquad \mbox{otherwise} 
      \end{array} \right.
      \end{equation}
where $D$ is the normalisation factor and $a$ is a quantity lower 
than or equal to half the nearest lattice spacing.
 \par 
Given this condition, two functions of the set referring to 
different sites on the lattice do not overlap and are therefore 
orthogonal. The usual second quantisation formalism can therefore 
be used without modification to deal with a system of identical 
bosons. The average value of the Hamiltonian is 
expressed with one- and two-body terms as follows:

      \begin{eqnarray}\label{e3}
        &< H > = \sum_i \frac{1}{ 2 \mu} \int 
          \varphi^{\ast}_i(\mbox{\boldmath $r$}) \,  \widehat {p^2 }
       \, \varphi_i(\mbox{\boldmath $r$}) \, d^3r \nonumber \\ &+ 
      {1 \over 2} \sum_i \sum_j
      \int\!\!\int \varphi^{\ast}_i(\mbox{\boldmath $r$}) \,
      \varphi_i(\mbox{\boldmath $r$}) \,
       v(|\mbox{\boldmath $r$}-\mbox{\boldmath $r'$}|) \, 
      \varphi^{\ast}_j(\mbox{\boldmath $r'$}) \,
      \varphi_j(\mbox{\boldmath $r'$}) \, d^3r \, d^3r'
      \end{eqnarray}

where  $\widehat {p^2} = \widehat{\mbox{\boldmath $p$}} \cdot 
     \widehat{\mbox{\boldmath $p$}}$.
  Here,  $\widehat{\mbox{\boldmath $p$}}$ is the one particle momentum 
      operator and $v(|\mbox{\boldmath $r$}-\mbox{\boldmath $r'$}|)$  
       is the two-body potential.
 \par
The expression for average potential energy can be re-written 
as
      \begin{equation}\label{e4}
     {1 \over 2} \sum_i \sum_{n^{(i)}} c_{n^{(i)}}
      \int\!\!\int \varphi^{\ast}_i(\mbox{\boldmath $r$})\, 
      \varphi_i(\mbox{\boldmath $r$}) \,
       v(|\mbox{\boldmath $r$}-\mbox{\boldmath $r'$}|) \, 
      \varphi^{\ast}_{n^{(i)}}(\mbox{\boldmath $r'$}) \,
      \varphi_{n^{(i)}}(\mbox{\boldmath $r'$}) \, d^3r \, d^3r'
      \end{equation}

The second sum is on the neighbouring $n^{(i)}$ of each site $i$, and 
 $c_{n^{(i)}}$ 
indicates the number  of the $n^{th}$ next nearest  . Since the realistic 
potential to be used goes rapidly to zero, it can be further 
simplified with accuracy by truncating the sum on the neighbours 
at a specific $\bar{n}$ and ignoring the surface effects for large $N$.
 \par 
The result for the average value of the Hamiltonian is then:

      \begin{eqnarray}\label{e5}
       &< H > =  \frac{N}{2 \mu} \int 
      \varphi^{\ast}_l(\mbox{\boldmath $r$}) \, 
        \widehat {p^2 } \, 
        \varphi_l(\mbox{\boldmath $r$}) \, d^3r + \nonumber \\ & 
       \frac{N}{2} 
        \sum_{n^{(l)}}^{\bar n} c_{n^{(l)}} 
      \int\!\!\int \varphi^{\ast}_l(\mbox{\boldmath $r$}) \,
       \varphi_l(\mbox{\boldmath $r$}) \,
       v(|\mbox{\boldmath $r$}-\mbox{\boldmath $r'$}|) \,
       \varphi^{\ast}_{n^{(l)}}(\mbox{\boldmath $r'$}) \,
      \varphi_{n^{(l)}}(\mbox{\boldmath $r'$}) \, d^3r \, d^3r'
      \end{eqnarray}

where the suffix $l$ indicates a generic lattice point. The one-body 
term (i.e. average kinetic energy) can be expressed analytically. A 
closed result for the double integral expressing the average 
potential energy is more difficult to obtain. However, this can be 
expressed by a single integration in the Fourier space, taking into
 account the isometric properties of the Fourier transform and using the 
convolution theorem for the quantity:
      \begin{equation}\label{e6}
      U(\mbox{\boldmath $r'$}) = \int v(|\mbox{\boldmath $r$}-
      \mbox{\boldmath $r'$}|)
      \, \varphi^{\ast}_{n^{(l)}}(\mbox{\boldmath $r'$})
       \,\varphi_{n^{(l)}}(\mbox{\boldmath $r'$}) \, d^3r' \qquad .
      \end{equation}    
Integration was then easily performed numerically.
 \par
The $N$ bosons were then considered as Krypton atoms. Thus, the 
corresponding value was used for mass $\mu$.
 As for two-body interaction, the two Yukawa potential was used as follows:
      \begin{equation}\label{e7}
      v(r)=-\epsilon \, b \, [ e^{- m (r/\sigma-1)} -
       e^{- n (r/\sigma-1)} ] \, / \,(r/\sigma)
      \end{equation}    
with $b = 2.026$, $m = 2.69$ and $n = 14.70$. The above expression is 
well known in scientific literature and is widely used for the best 
possible reproduction of the Van der Waals potential in the case of 
noble gases \cite{pol}. In the case of Krypton, $\epsilon = 170\,$~ K
 and $\sigma = 3.6\,$~\AA. As 
can be seen in Eq.~(\ref{e5}), the average energy is additive, as is to be 
expected. This energy depends on the type of lattice, its nearest 
spacing $d$ and parameters $\lambda$ and $a$ relative to the function
 $\varphi(\mbox{\boldmath $r$})$.
 \par 
After some work on the computer, the minimum value for 
average energy was obtained in the case of a face centred cubic 
lattice with the nearest neighbour distance $d = 3.953 \,$~\AA\ and when 
$\lambda = 91.33/\sigma $. Parameter $a$ was fixed at the beginning as equal
 to half of $d$. However,  suitable values of $\lambda$ are so large and
 $\varphi(\mbox{\boldmath $r$})$
therefore decreases so rapidly (in accordance with the initial 
assumption of non-overlapping) that $a$ can safely be left to go on 
to infinity.
 \par
Once the probe function was characterised, it was easy to 
deduce both cohesive energy $U$ and bulk modulus $B$. The results 
obtained were: $U = -2690 \,$ cal/mole and $B = 33.4 \,$ kbar. The 
experimental values for solid Krypton at 0 K, which actually 
crystallises as a face centred cubic, are indicated \cite{bar} as
 $d = 3.992 \,$ \AA,  $U = -2666 \,$ cal/mole and $B = 34.3 \,$ kbar.
 Although it was not the 
aim here to make a solid state computation, comparison of the 
results with the experimental ones suggests that the probe state 
specified above represents a combination of low-lying states.
 \par
A function  $\varphi(\mbox{\boldmath $r$})$ centred on the ith
 reticular point can, of 
course, be associated with more than one of the spinless bosons. 
To see how this can affect the average energy, the following 
quantity must be considered:

      \begin{equation}\label{ea}
       W =         \int\!\!\int \varphi^{\ast}_i(\mbox{\boldmath $r$})\, 
      \varphi_i(\mbox{\boldmath $r$}) \,
       v(|\mbox{\boldmath $r$}-\mbox{\boldmath $r'$}|) \, 
      \varphi^{\ast}_i(\mbox{\boldmath $r'$}) \,
      \varphi_i(\mbox{\boldmath $r'$}) \, d^3r \, d^3r' \qquad .
      \end{equation}

 \par
If $p$ particles occupy the same lattice site, the contribution to 
the average potential energy from their interaction is $ p(p - 1) W/2$. 
In the case of a potential with a short-range repulsive part, this 
contribution is presumably positive. In the present case, with the 
indicated values of $\lambda$ and $d$, $W$ is not only positive, but also ten 
million times greater than the absolute value of the (negative) 
interaction energy per particle. It can be concluded that the lowest 
value for the average energy of the system is reached in the case 
considered of only one particle per site. This conclusion does not 
change in the case of bosons with non-zero spin.

 \vskip .9 true cm
 \noindent{\bf 4. THE SPREAD OF THE CENTER OF MASS POSITION}
 \vskip .6 true cm
 \par
Using the probe wave function obtained, we can calculate the 
spread of the center of mass position. Components
 $R_i$  and $P_i$  of the 
center of mass position vector:
      \begin{equation}\label{e8}
      \mbox{\boldmath $R$} = \sum \mbox{\boldmath $r$}_i / N
      \end{equation}    
and of the conjugate momentum:
      \begin{equation}\label{e9}
      \mbox{\boldmath $P$} = \sum \mbox{\boldmath $p$}_i
      \end{equation}    
have an average value of zero. So the requirement on the mean
value of the total momentum is satisfied. The average squares of
$R_i$ are:

      \begin{equation}\label{e10}
      < R_i^2 > = {4 \over \lambda^2 N }     \qquad  (i=1, 2, 3)
      \qquad .
      \end{equation}    
The mean square deviations:
      \begin{equation}\label{e11}
      \chi_i = < R_i^2 > -  < R_i >^2  = {4 \over \lambda^2 N }
       \qquad  (i=1, 2, 3)
      \end{equation}    
therefore depend on $N$ as $1/N$. The following is obtained for the 
mean-square deviations of the momentum components:
      \begin{equation}\label{e12}
      \omega_i = < P_i^2 > -  < P_i >^2 = \hbar^2 \lambda^2 N / 12
           \qquad  (i=1, 2, 3) \qquad .
      \end{equation}    
These are proportional to $N$ and it can be seen that the 
indeterminacy product is very close to the minimum value.
 \par
It can be observed that the mean-square deviations for the 
center of mass velocity components behave like $1/N$. In other 
words, both of the collective (pseudo)conjugate variables 
represented by the center of mass coordinate and its velocity have 
vanishing small dispersion in the limit of large $N$.
 \par 
We have concentrated so far on the properties of the 
state vector obtained, which will be indicated as
      $\Phi(\mbox{\boldmath $r$}_1,\mbox{\boldmath $r$}_2, 
      \cdots)$.
However, since the origin and orientation of the coordinates are 
arbitrary, the states which can be obtained from it by translation 
and rotation must also be considered. The Hamiltonian is 
translationally and rotationally invariant, so these states have
 the same average energy. It is particularly interesting to investigate
 a linear superposition of them. For the sake of simplicity, we have 
 considered only a superposition of translated states:
      \begin{eqnarray}\label{eb}
       \Psi(\mbox{\boldmath $r$}_1,\mbox{\boldmath $r$}_2, \cdots) = 
        c_0 \, \Phi(\mbox{\boldmath $r$}_1,\mbox{\boldmath $r$}_2, \cdots)+
        c_1 \, \Phi(\mbox{\boldmath $r$}_1+\mbox{\boldmath $a$}_1,
      \mbox{\boldmath $r$}_2+\mbox{\boldmath $a$}_1,+ \cdots)+ \cdots 
       \nonumber \\  \cdots
      +  c_n \, \Phi(\mbox{\boldmath $r$}_1+\mbox{\boldmath $a$}_n+
      \mbox{\boldmath $r$}_2+\mbox{\boldmath $a$}_n, \cdots) \ .
       \end{eqnarray}
 \par
The above wave vector is still a completely symmetrised state 
of the boson system. Let us consider the very simple case in which 
all the $\Phi$ in the second member are centred so far apart 
from each other that none of them overlaps another. This is 
possible due to the way in which the functions
 $\varphi_i(\mbox{\boldmath $r$})$
 appearing in 
the product  (Eq.~(\ref{e1})) 
 defining the trial vector were chosen. In 
these conditions, the cross terms do not contribute to average 
energy or to average total momentum. Furthermore, the scalar product 
of each pair of $\Phi$ centred at different points is zero and the 
normalisation condition requires:
      \begin{equation}\label{ed}
         \sum_{i=0}^n |c_i|^2 = 1  \qquad .
      \end{equation}
 \par           
So the linear superposition considered has the same average energy 
as the starting state and the average value of the center of mass 
 momentum is still zero.
 The $\Psi(\mbox{\boldmath $r$}_1,\mbox{\boldmath $r$}_2, \cdots)$
 corresponds to the entire blob of solid Krypton being in
 a spatial superposition of 
localised states, which can be separated by any distance, however 
great. The resulting spread for the center of mass position is 
therefore unlimited, which unfortunately means that our 
energy considerations cannot be used to select only localised states.
 \par
It can be observed that, under the present conditions, the matrix element 
between two different $\Phi$ is zero not only in the case of 
energy and total momentum but also for any observable  
depending on the coordinates and momenta of the particles. This 
is the basic requisite of superselection rules for spatial
 macrosuperpositions of our localised states.
 However, it is better not to 
insist on this, since everything depends on having considered the 
initial functions $\varphi$ as equal to zero outside a finite spatial region.

 \vskip .9 true cm
 \noindent{ \bf 5. THE SELF GRAVITATING PARTICLES}
 \vskip .6 true cm
 \par
The system examined here is unlikely to represent a generic 
macroscopic body, even though it may be thought that interaction 
between particles with both a short-range repulsive part and a 
long-range attractive one is an essential characteristic 
as long as atoms or molecules of the same type can 
be considered as elementary constituents of a large homogeneous object.
 \par 
One way to generalise our model of a body is to consider the 
case in which the elementary constituents are identical fermions 
with spin $ s $. A lattice site may then be occupied by up to
 $ g = 2s + 1 $ 
particles with different spin states. But reference can be made to our 
previous considerations on this type of configuration: it was seen 
that the contribution to potential average energy is positive and very 
large when compared with the absolute value of (negative) 
average energy per particle in the case of one particle per site.  
Looking for a superposition of low-lying energy states, as long 
as we consider a spin independent two-body interaction of Van
 der Waals type, it does not matter whether the constituents of the body are 
bosons or fermions. The localised "natural states" being sought 
correspond to only one particle per site, whether a boson or a 
fermion, irrespective of the spin state the particle assumes there. 
In the present model, the spread of the center of mass position for 
a localised state with low average energy is therefore the same for 
bosons and fermions.
 \par
This independence from the type of constituent particles is 
presumably due to the type of potential used, with a deep 
minimum at finite distance. To see whether the situation changes 
for other two-body potentials, let us examine a system of identical 
self gravitating particles. The resulting model of a macroscopic 
body is not very realistic, but it may throw light on the point in 
question. The pair wise interaction is now:
           \begin{equation}\label{eg}
         v(|\mbox{\boldmath $r$}-\mbox{\boldmath $r'$}|) =
      - \frac {\kappa}{|\mbox{\boldmath $r$}-\mbox{\boldmath $r'$}|}
      \qquad .
      \end{equation}
 \par
Let us first consider the case of identical spinless bosons. When 
searching for a localised state of very low average energy, all 
particles should preferably be in the same spatial state, as has 
been seen in the case of the (relativistic) collapse of a boson star 
\cite{car}. The following $N$-body wave function (which was useful in the 
case of boson stars) is used as the trial state vector:
      \begin{equation}\label{eh}
      \Theta(\mbox{\boldmath $r$}_1,\mbox{\boldmath $r$}_2, \cdots)=
      D(\beta) \, \prod^N_{i=1} \, e^{-\beta r_i}
      \end{equation}
where  $D(\beta)$ is the normalisation factor. Corresponding average 
energy is given by:
            \begin{equation}\label{ei}
      <E>=N \, \hbar^2 \beta^2/(2\mu)-5 \kappa N(N-1)\beta/16
      \end{equation}
where $\mu$ is the mass and $N$ the number of particles. The 
minimum value is reached when:
      \begin{equation}\label{el}
      \beta=5 \,\kappa \, \mu (N-1)/(16 \,\hbar^2) \qquad .
      \end{equation}
 \par
Now the average energy is no longer proportional to $N$, a result already 
known \cite{lev2}. The following average values for the center of mass 
coordinates are obtained:
            \begin{equation}\label{em}
      < R_i> = 0     \qquad  (i=1, 2, 3)
      \end{equation}    

      \begin{equation}\label{en}
      < R_i^2 > =  g^2 /( N(N-1)^2)     \qquad  (i=1, 2, 3)
      \end{equation}    
where $g = 16 \, \hbar^2/(5 \kappa \, \mu) $. The mean-square deviations of the 
center of mass position components depend on $N$ as $1/N^3$. The 
mean-square deviations of the components of momentum are 
proportional to $N(N - 1)^2$, while the indeterminacy product is 
equal to $\hbar/\sqrt{3}$.
 \par
If the self gravitating particles are fermions, an anti-
symmetrised wave function must be used as a trial state vector. 
In this case, the calculations would be complicated and it is 
therefore preferable to resort to semi-classical 
approximation, finding the minimum of the Thomas-Fermi energy 
functional:

            \begin{equation}\label{ep}
      \int \tau(\mbox{\boldmath $r$}) \, d^3r-\frac{\kappa}{2} \int\!\!\int
      \frac{ \rho(\mbox{\boldmath $r$})\, \rho(\mbox{\boldmath $r'$})}
      {|\mbox{\boldmath $r$}-\mbox{\boldmath $r'$}| } \, d^3r \, d^3r'
      \end{equation}    
with
      \begin{equation}\label{eq}
       \tau(\mbox{\boldmath $r$}) = \frac{ e \, \hbar^2}{q^{2/3} \, \mu}\, 
      \rho^{5/3}(\mbox{\boldmath $r$})
      \end{equation}    
where $e \simeq 5 $ is a numerical factor, 
 $\rho(\mbox{\boldmath $r$})$
 is the spatial distribution 
of the particles and $q$ is the occupation number.
 \par
 We assumed
the following trial spatial distribution normalised to the number $N$ of 
fermions:
      \begin{equation}\label{er}
       \rho(\mbox{\boldmath $r$}) =N \, \gamma^3 e^{-\gamma r}/(8\pi)
      \end{equation}
where $\gamma$ is a parameter to be varied. When considering particles 
with 1/2 spin $(q = 2)$, it is found that the minimum of 
 Eq.~(\ref{ep})  (i.e. 
of average energy) is reached when
  $\gamma = f \kappa \mu N^{1/3}/\hbar^2 $,
 $f$ being a numerical factor of  order unity.
With the aid of the corresponding spatial particle distribution, 
the mean-square deviations for the center of mass position 
components can be calculated. The following is obtained:

      \begin{equation}\label{es}
      \chi_i=\frac{4 \, \hbar^2}{b^2 \, \kappa^2 \, \mu^2 \, N^{5/3}}           
       \qquad  (i=1, 2, 3) \qquad .
      \end{equation}
It can be seen that fermions and bosons behave differently as regard
the spread of the centre of mass position.
 In both cases, however, the position spread
 decreases as $N$ increases.
 \vskip .9 true cm
 \noindent{\bf 6. FINAL REMARKS}
 \vskip .6 true cm
 \par
A brief summary of the findings in the case of our model of 
macroscopic body is needed here. For the states identified using 
our trial wave function and our energy considerations, the average total 
momentum is zero and the center of mass position is well 
localised. A Galilean transformation \cite{lev} can also be applied to 
provide the body with a required average momentum (velocity).
The state considered can also be slightly modified, for example 
by changing the value of $\lambda$ for some or all of the
functions  $\varphi(\mbox{\boldmath $r$})$
centred on the lattice points or by allowing some other points of 
the lattice to be occupied and some vacancies to appear. By 
applying a translation and a Galilean transformation, it is then 
possible to have various states of an isolated body which 
correspond to a well defined position and velocity of the center of 
mass.
 \par
These states correspond to a nearly minimum-uncertainty state 
for the center of mass, and the mean-square deviations of the 
center of mass position components decrease as $1/N$ as the number  
 $N$ of constituents increases. The above property does not seem 
to differ whether the constituent particles are bosons or fermions.
Unfortunately, there are a large number of states which have 
the same low average energy and are not at all localised. These 
are given by the macrosuperpositions of the previous states 
centred in different places and with different orientations.
 Apart from this, since the body is isolated, the spread of 
the center of mass position increases in any case with the passing 
of time as $t^2$. To obtain a classical behaviour in the 
macroscopic limit, it is evidently necessary to resort to mechanisms such as 
decoherence caused by the external environment. We have 
obtained macroscopic states which are basically eigenstates of 
position ( although these constitute a set which has measure zero
 compared to all other delocalised states with the same energy ), starting 
with considerations regarding the inevitable interaction of the body 
with surrounding objects. This at least seems to indicate that we 
were moving in the right direction.
 \par
Since the criterion adopted was not sufficient to select only 
localised states, we asked ourselves what further requirement was 
needed to reach this objective. It may be conjectured that besides 
being a superposition of low-lying energy states, the macroscopic 
states must be a product of the wave functions of the 
individual constituents. However, it is not 
clear how this condition can be justified.
 \par
The case of a system of $N$ self gravitating particles was then 
investigated. For those localised trial states which satisfy our 
energy criterion and in the case of bosons, the mean-square
 deviations of the center of mass position components decrease as $1/N^3$
 and the indeterminacy product is close to the minimum value. In the case 
of fermions, the center of mass is still well localised but the mean-
square deviations of its position components decrease as $1/N^{5/3}$.
 \par
In conclusion we have not, of course, answered the big question about the 
special role which position seems to play in the macroscopic limit 
of quantum mechanics. However, there does seem to be some 
merit in demonstrating, within our solid noble gas model, the plausibility
 of well localised macroscopic bodies corresponding to a minimum 
uncertainty product for their center of mass. 
 \par
 Of course we cannot consider to be generally valid the result we
 obtained in this connection by a trial wavefunction which is
 not a very generic wavefunction of the desidered kind. 
  It was selected mainly on grounds of 
computational convenience and having in mind the classical ground
 state. But since it would be very energetically expensive to have
 any inter-particle separation differing greatly from the classical
 arrangement, a " natural state " of the system considered cannot have
 any of its constituent particles in a flagrant macrosuperposition,
 except those that reflect the symmetries of the Hamiltonian,
 in our case global translation, rotation and reflection of all
 particles together. These considerations seems indicate that the
 localised state of the system cannot differ greatly from our
 trial wave function if its energy is to be close to minimal.
 So the conclusion that states with uncertainty product close
 to minimal are in a sense singled out by energy considerations
 may be generally correct.
 \par
It interesting 
to note at last that the study of the apparent collapse of the wave 
function caused by scattering \cite{teg} seems to indicate that 
macroscopic bodies tend to be characterised by a minimum 
packet.

       \vskip 1.4 true cm

      \end{document}